\documentclass[twocolumn, secnumarabic, amssymb, amsmath,nobibnotes, aps, prl]{revtex4}

\begin{document}

\title{Thermodynamics of An Ideal Generalized Gas: II Means of Order $\alpha$} 
\author{B. H. Lavenda}
\email{bernard.lavenda@unicam.it}
\affiliation{Universit\'a degli Studi, Camerino 62032 (MC) Italy}
\date{\today}
\newcommand{\sumi}{\sum_{i=1}^{n}\,}
\newcommand{\sumj}{\sum_{j=1}^{n}\,}
\newcommand{\sumij}{\sum_{i,j=1}^{n}\,}
\newcommand{\sumk}{\sum_{i=1}^{k}\,}
\newcommand{\sumgr}{\sum_{i=1}^{>}\,}
\newcommand{\sumle}{\sum_{i=1}^{\le}\,}
\newcommand{\prodi}{\prod_{i=1}^n\,}
\newcommand{\half}{\mbox{\small{$\frac{1}{2}$}}}
\newcommand{\fourth}{\mbox{\small{$\frac{1}{4}$}}}
\newcommand{\third}{\mbox{\small{$\frac{1}{3}$}}}\newcommand{\twothirds}{\mbox{\small{$\frac{2}{3}$}}}

\begin{abstract}
 The property that power means are monotonically increasing functions of their order is shown to be the basis of the second laws not only for processes involving heat conduction but also for processes involving deformations. In an $L$-potentail equilibration the final state will be one of maximum entropy, while in an entropy equilibrium the final state will be one of minimum $L$. A metric space is connected with the power means, and the distance between means of different order is related to the Carnot efficiency. In the ideal classical gas limit, the average change in the entropy is shown to be proportional to the difference between the Shannon and R\'enyi entropies for nonextensive systems that are multifractal in nature. The $L$-potential, like the internal energy, is a Schur convex function of the empirical temperature, which satisfies Jensen's inequality, and serves as a measure of the tendency to uniformity in processes involving pure thermal conduction.
\end{abstract}
\maketitle
\section{Mathematical versus physical inequalities}
Since the harmonic-arithmetic-geometric mean inequalities are  particular manifestations of the property that power means are increasing functions of their order, it was thought that the second law inequality could be derived from this property. This would avoid having to resort to experiment to determine the sign of the entropic change in processes involving non-quasi static changes.\par
Apart from an exercise that can be found in Sommerfeld's book \emph{Thermodynamics and Statistical Physics\/} \cite{Sommerfeld}, where he shows that the increase in entropy of an ideal classical gas (ICG) which has come to equilibrium is \lq\lq a generalization of the inequality between arithmetical and geometrical means\rq\rq, the real impetus began with a series of papers in the early '80's by Landsberg and co-worker \cite{PTL-bis} to generalize the arithmetic-geometric mean inequality to cases of negative heat capacities. Sidhu \cite{Sidhu} generalized their results to arbitrary power means of order greater than one, and excluded negative powers on the basis that they would be in conflict with the third law.\par
Yet all these results could be found in an earlier paper by Cashwell and Everett \cite{Cashwell} who, because they dealt with temperature dependent heat capacities, were dealing with an ideal generalized gas (IGG). Their work was, however, limited to processes of pure thermal conduction in which a system was divided into a number of cells whose initial temperatures were all not equal and subsequently allowed to interact thermally by replacing the adiabatic partitions by diathermal ones. The mass fractions played the role of a complete probability distribution. Probabilistic notions naturally arise when the uncontrollable processes concerning heat transfer among the cells occur. The final equilibrium state was characterized by a common mean temperature determined from the conservation of the internal energy. The second law followed from the property that the final common temperature was greater than that which would have been obtained in an entropy-conserving equilibration, or that the power means are monotonically increasing functions of their order.\par
It was realized that in an entropy-conserving equilibration \cite{PTL}  a final common temperature would be reached that would be lower than an energy-conserving equilibration, implying that energy has been extracted from the $n$-body system for the performance of external work. This would then provoke a negative change in the internal energy making it comparable to an entropy evolution criterion. It would thus appear that the first and second laws have exchanged roles. The problem inherent to such a formulation, apart from processes involving pure thermal conduction, is that the power means determined from an energy- or entropy-conserving equilibration are not comparable: the temperature dependencies are different but the volume dependencies are not, since both internal energy and entropy are first-order homogeneous functions of the volume. Thus, for  processes other than pure heat conduction, the two laws are not comparable. Moreover, in the case of pure deformations, there would be no evolution criterion at all because both potentials are first-order homogeneous functions of the volume. \par
For processes involving pure thermal conduction, a formulation different from the Cashwell-Everett one was given by Abriata \cite{Abriata}. In that formulation, an increase in entropy, on the average, is associated with going from a less uniform to a more uniform temperature distribution, or becoming \lq less spread out\rq. As such it can be formulated as a problem of majorization \cite{Marshall}, with the internal energy, or the $L$-potential, playing the role of a Schur convex function of the emprical temperature. Although the entropy need not be a Schur concave function of the empirical temperature, it must necessarily be one of the energy. In this way a continuous temperature distribution involving the Schur convex function of the energy is contrasted to a possibly discrete energy distribution related to the Schur concave function of the entropy. Majorization is considered in the last section.
\section{Comparable means and the laws of thermodynamics}
The conventional forms of the first and second laws are only comparable for processes involving pure heat conduction. Thus, Cashwell and Everett's results \cite{Cashwell} constitute a particular case where the internal energy, $E$, and the $L$-potential differ by a constant factor since the volume is held constant. Their results cannot be generalized to processes involving work.\par
Means are said to be comparable if there exists an inequality \cite{HLP}
\begin{eqnarray}
\lefteqn{\mathfrak{M}_{f}(z)=f^{-1}\left(\sumi p_if(z_i)\right)}\nonumber\\
& & \le
g^{-1}\left(\sumi p_ig(z_i)\right)=\mathfrak{M}_{g}(z) \label{eq:mono}
\end{eqnarray}
between them, where $\{p_i\}$ is a complete probability distribution, and $f^{-1}(z)$ the inverse function. A well-known necessary and sufficient condition for inequality (\ref{eq:mono}) to hold is that the composite function $g\circ f^{-1}$ be convex on the interval $[z_c,z_h]$,  if $g$ is increasing \cite{HLP}. Inequality (\ref{eq:mono}) is satisfied by power  means, where the generators are $g=z^{\alpha}$ and $f=z^{\alpha-1}$.\par Since the concept of absolute entropy has no meaning in thermodynamics,  one cannot distinguish between the Nernst and Planck formulations of the third law \cite{Einbinder}. This is translated into the property of equivalent means \cite{HLP}: in order for
\[\mathfrak{M}_{F}(z)=\mathfrak{M}_{f}(z),\]
it is both necessary and sufficient that
\[F=a\,f+b,\]
where $a$ and $b$ are constants, and $a\neq0$.\par
Consider a system comprised of $n$ cells, adiabiatically isolated from the environment. Initially the walls of the cells are rigid and adiabatic. When the walls are replaced by deformable, diathermal ones, the initial probability that the $i$th cell will have a linear dimension, $R_i=V_i^{1/(q-r)}$, and an empirical temperature, $t_i=T^{1/r}$, is $p_i$. Probabilities enter naturally when dealing with processes of heat exchange:  Heat  is the uncontrollable form of work \cite{JJ}, and temperature is its measure.\par
If $\vec{z}$ is an $n$-tuple of initial values of the adiabatic variables, $z_i=\left(t_iR_i\right)^r$, and $\vec{\zeta}$ is an $n$-tuple of their final, equilibrium, values  then the average changes in $L$ and $S$ when the subsystems have been placed into thermal and mechanical contact are 
\begin{equation}
\overline{\Delta L}=\sumi p_i\int_{z_i}^{\zeta_i}\,dL(z)=
\mathfrak{M}_{\alpha}^{\alpha}(\zeta)-\mathfrak{M}_{\alpha}^{\alpha}(z), \label{eq:L-law}
\end{equation}
and
\begin{eqnarray}
\overline{\Delta S} & = & \sumi p_i\int_{z_i}^{\zeta_i}\,
\frac{dL(z)}{z}\nonumber\\ & = &  \frac{\alpha}{\alpha-1}\left\{\mathfrak{M}_{\alpha-1}^{\alpha-1}(\zeta)-\mathfrak{M}_{\alpha-1}^{\alpha-1}(z)\right\}. \label{eq:S-law}
\end{eqnarray}
The vanishing of either relation, (\ref{eq:L-law}) or (\ref{eq:S-law}) would determine a uniform mean value of $\zeta$. From an $L$-equilibration  it would be $\mathfrak{M}_{\alpha}(z)$, while from an $S$-equilibration the final mean value would be $\mathfrak{M}_{\alpha-1}(z)$. On the strength of (\ref{eq:mono}), the former would be larger than the latter.
\par
Consider an $L$-equilibrating transition where (\ref{eq:L-law}) vanishes. Then, if we divide the cells into two groups those for which $z_i\le\mathfrak{M}_{\alpha}(z)$, marked by a \lq\lq $\le$ \rq\rq\ on the upper limit of the summation sign, and those for which $z_i>\mathfrak{M}_{\alpha}(z)$, indicated by a \lq\lq $>$ \rq\rq\ on the summation sign, we find 
\begin{eqnarray*}
\lefteqn{\sumle p_i\int_{z_i}^{\mathfrak{M}_{\alpha}(z)}\,\frac{dL(z)}{z}  > 
\frac{1}{\mathfrak{M}_{\alpha}(z)}\sumle p_i\int_{z_i}^{\mathfrak{M}_{\alpha}(z)}\,dL(z)}\\
& = & \frac{1}{\mathfrak{M}_{\alpha}(z)}\sumgr p_i\int_{\mathfrak{M}_{\alpha}(z)}^{z_i}\,dL(z)
>\sumgr p_i\int_{\mathfrak{M}_{\alpha}(z)}^{z_i}\,\frac{dL(z)}{z}.
\end{eqnarray*}
\par
Since $\overline{\Delta S}>0$  for $\overline{\Delta L}=0$, no matter what the initial $n$-tuple of $z$ values are, the inequality  \cite{Cashwell-bis}
\begin{equation}
\sumi p_i\int_{z_i}^{\zeta_i}\,\frac{dL(z)}{z}\le\sumi p_i\int_{z_i}^{\mathfrak{M}_{\alpha}(z)}\,\frac{dL(z)}{z} \label{eq:II}
\end{equation}
is a consequence of  the fact that the last term in
\begin{eqnarray*}
\lefteqn{\sumi p_i\int_{z_i}^{\zeta_i}\,\frac{dL(z)}{z}}\\
& = & \sumi p_i\int_{z_i}^{\mathfrak{M}_{\alpha}(z)}\,\frac{dL(z)}{z}-\sumi p_i\int_{\zeta_i}^{\mathfrak{M}_{\alpha}(z)}\,\frac{dL(z)}{z}
\end{eqnarray*}
is positive.\par Hence, the average entropy change is greatest when the final state has a uniform mean $\mathfrak{M}_{\alpha}(z)$. In the particular case of pure thermal convection this result can be found in Cashwell and Everett \cite{Cashwell}. However, their result cannot be generalized to more general processes involving deformations because  the average change in the internal energy, derived from the first law, is not comparable to the average change in entropy.\par
The counterpart of  maximum entropy in the state of uniform mean $\mathfrak{M}_{\alpha}(z)$ in an $L$-equilbration is  minimum  $L$ in a state of a uniform mean $\mathfrak{M}_{\alpha-1}$ that results in an $S$-equilibration. In an $S$-equilibrating transition
\begin{eqnarray*}
\frac{1}{\mathfrak{M}_{\alpha-1}(z)}& \sumle & p_i\int_{z_i}^{\mathfrak{M}_{\alpha-1}(z)}\,dL(z)\\ & < & \sumi p_i\int_{z_i}^{\mathfrak{M}_{\alpha-1}(z)}\,\frac{dL(z)}{z}\\
& = & \sumgr p_i\int_{\mathfrak{M}_{\alpha-1}(z)}^{z_i}\,\frac{dL(z)}{z}\\
& < & \frac{1}{\mathfrak{M}_{\alpha-1}(z)}\sumgr p_i\int_{\mathfrak{M}_{\alpha-1}(z)}^{z_i}dL(z),
\end{eqnarray*}
implying that $\overline{\Delta L}<0$ when $\overline{\Delta S}=0$. The minimum property follows from the fact that the last term in
\begin{eqnarray*}
\lefteqn{\sumi p_i\int_{z_i}^{\zeta_i}\,dL(z)=}\\
& & \sumi p_i\int_{z_i}^{\mathfrak{M}_{\alpha-1}(z)}\,dL(z)-\sumi p_i\int_{\zeta_i}^{\mathfrak{M}_{\alpha-1}(z)}\,dL(z)
\end{eqnarray*}
is always negative, regardless of the initial $n$-tuple $\vec{z}$.\par Hence, the mean $\mathfrak{M}_{\alpha-1}(z)$ minimizes $L$. Although the $L$-potential is nonextensive, it nevertheless  shares properties in common to a free energy in manifesting a spontaneous tendency to decrease in the presence of irreversible processes. We shall return to this in the final section of this paper.
\section{Bounds on Mean Temperatures and Volumes}
Carath\'eodory's principle states that there are always neighboring states to any given state that are inaccessible to it by an adiabatic process, whether it be quasi-static or not \cite{Caratheodory}. This guarantees that there are sets of surfaces $\psi_i(z)=\mbox{const}$., where $\psi_i(z)$ is either $S(z)$ or $L(z)$, that do not intersect with each other. However, it does not say what those states are; this must come from another principle.\par This principle states that those states which are adiabatically accessible from any given state must increase the average internal energy \cite{Buchdahl}, i.e.,
\begin{eqnarray}
\overline{\Delta E}& = & L(z)\sumi p_i\int_{V_i}^{V_f}dV^{-s}\nonumber\\
& = & L(z)\left\{V_f^{-s}-\sumi p_iV_i^{-s}
\right\}\ge0 \label{eq:E-L}
\end{eqnarray}
or, equivalently,
\begin{eqnarray}
\overline{\Delta E} & = & S(z)\sumi p_i\int_{t_i}^{t_f}dt^r\nonumber\\
& = & S(z)\left\{t_f^r-\sumi p_it_i^r\right\}\ge0 \label{eq:E-S},
\end{eqnarray}
since they are linked by the adiabatic constraints $z_i=\mbox{const}$., $\forall i$.\par
According to (\ref{eq:E-L}), an adiabatic transition to a state with a final volume, $V_f$, is possible so long as there is an increase in the volume. In the limit as $s\rightarrow\infty$, i.e., $q\rightarrow r$,
\[
V_f\ge\lim_{s\rightarrow\infty}\left(\sumi p_iV_i^{-s}\right)^{-1/s}\rightarrow V_{\min}, \]
where $V_{\min}$ is the  volume of the smallest cell. Thus, the larger the adiabatic index, $s$, the more weight is given to cells of smaller size. And since the process is adiabatic, inequality (\ref{eq:E-S}) says that the final temperature 
\[t_f\ge\left(\sumi p_it_i^r\right)^{1/r},\]
cannot be inferior to the minimum mean temperature, $\mathfrak{M}_r(t)$.\par
By contrast, in a process where all the cells have a common empirical temperature, $t$, the maximum work that is performed on the system is one where the final volume, $V_f$, satisfies
\begin{eqnarray*}\overline{\Delta L} & = & cT^{\alpha}\sumi p_i\int_{V_i}^{V_f}dV^{\alpha s}\\
& = & cT^{\alpha}\left\{V_f^{\alpha s}-\sumi p_iV_i^{\alpha s}\right\}\le0,
\end{eqnarray*}
since in the limit $s\rightarrow\infty$
\[V_f\le\left(\sumi p_iV_i^{\alpha s}\right)^{1/\alpha s}\rightarrow V_{\max}.\]
The upper limit to the final mean volume is the volume of the largest cell, $V_{\max}$.\par
Finally, in a process where no work is performed, and  the cells all have the same volume, $V_i=V$, 
\[\overline{\Delta L}=cV^{\alpha s}\sumi p_i\int_{t_i}^{t_f}dt^q=cV^{\alpha s}\left\{t_f^{q}
-\sumi p_it_i^q\right\}\le0.\]
Since only processes of pure heat conduction are involved, the average energy change manifests the same trend to decrease. The highest attainable temperature is one where there is an $L$-conserving equilibration, and
\[t_f=\left(\sumi p_it_i^q\right)^{1/q}.\]
The largest mean temperature $\mathfrak{M}_q(t)$ is proportional to the internal energy, and the difference between this mean temperature and the minimum mean temperature $\mathfrak{M}_r(t)$ is related to the system's capability of performing work, as we shall now discuss.
 \section{Metric space of power mean differences}
 It has long been appreciated that thermodynamic surfaces of convex energy or concave entropy lack the important topological element of a metric.  There is nothing in classical thermodynamics that would play the role of a distance function and would satisfy the triangle inequality.\par
However, the absolute difference of power means have been shown to represent a distance, or metric, on the set of all continuous and monotonic functions in the domain $[z_c,z_h]$ \cite{Cargo}. \par
 Consider the average change in the $L$-potential
 \begin{equation}
 \overline{\Delta L}=c\left(\sumi p_iz_i^{\alpha-1}\right)^{\alpha/(\alpha-1)}-c\sumi p_iz_i^{\alpha},
 \label{eq:L-bis}
 \end{equation}
 where the first term is the result of an $S$-equilibration. Setting $x_i=z_i^{\alpha}$, (\ref{eq:L-bis}) can be written as the difference of two power means
 \[
 \overline{\Delta L}=c\left(\sumi p_i x_i^{\gamma}\right)^{1/\gamma}-c\sumi p_ix_i<0,
 \]
 where $\gamma=(\alpha-1)/\alpha$, and the inequality follows from (\ref{eq:mono}).\par Alternatively, in an $L$-equilibration, the average entropy increases by an amount
 \[\overline{\Delta S}=\frac{c}{\gamma}\left\{\left(\sumi p_iy_i^{1/\gamma}\right)^{\gamma}-\sumi p_iy_i\right\}>0,\]
for the same reason.
 \par
 The distance $d(f,g)$ between the generators $f$ and $g$ of the power means is defined by
 \begin{equation}
 d(f,g):=\sup_{z}\left\{\left|\mathfrak{M}_f(z)-\mathfrak{M}_g(z)\right|:z_c\le z\le z_h\right\}.\label{eq:d}
 \end{equation}
 For pure thermal conduction, the inequality
 \[\left|\mathfrak{M}_f(z)-\mathfrak{M}_g(z)\right|\le z_h-z_c,\]
 shows that the maximum distance between $f$ and $g$ is bounded by the Carnot efficiency
 \begin{equation}
 \frac{\left|\mathfrak{M}_f(z)-\mathfrak{M}_g(z)\right|}{z_h}\le\eta_t. \label{eq:eta}
 \end{equation}
 All other inequalities that we will derive will be sharper, and, consequently, less efficient.\par
 The distance (\ref{eq:d}) is obviously symmetric, and  satisfies the triangle inequality
  \begin{eqnarray}
 d(f,g)& = &\left|\sumi p_ix_i-\left(\sumi p_ix_i^{\gamma}\right)^{1/\gamma}\right|\nonumber\\
 &\le & \left|\sumi p_ix_i-\sumi p_ix_i^{\gamma}\right|\nonumber\\
 & + & \left|\sumi p_ix_i^{\gamma}-\left(\sumi p_ix_i^{\gamma}\right)^{1/\gamma}\right|.
 \label{eq:tri}
 \end{eqnarray}
 If $f$ denotes the generator of the weighted arithmetic mean and $g$ the generator of the mean of order $\gamma$, then
 \begin{eqnarray*}
\lefteqn{\left|\sumi p_ix_i-\sumi p_ix_i^{\gamma}\right|}\\
&  &\le \sumi p_i\|f-g\|=\|f-g\|,\end{eqnarray*}
where $\|\cdot\|$ denotes the norm,
\[\|f-g\|=\sup_{x}\left\{|f(x)-g(x)|:x_c\le x\le x_h\right\}.\] Likewise, setting $w_i=x_i^{\gamma}$, the second term in (\ref{eq:tri}) is bounded from above by
\begin{eqnarray*}
\lefteqn{\left|\sumi p_iw_i-\left(\sumi p_iw_i\right)^{1/\gamma}\right|}\\
& & \le
\left|\sumi p_iw_i-\sumi p_iw_i^{\gamma}\right|\le\sumi p_i\|f-g\|=\|f-g\|,
\end{eqnarray*}
since $\mathfrak{M}_{\gamma}(w)\le\mathfrak{M}_1(w)$ for $\gamma<1$. The equality sign pertains to the case where all the $w_i$ are equal. \par
This proves that
\begin{equation}
\left|\overline{\Delta L}\right|\le 2c\left\|f-g\right\|,\label{eq:L-ineq}
\end{equation}
and the distance is induced by the norm. The properties of the metric space of equivalent classes on the set of all continuous and montonic functions have been elucidated in \cite{Cargo}. In particular, the metric space is separable since there is a countable subset everywhere dense in it, like that of the real line.\par
In some cases it is possible to derive analytic inequalities which are sharper than (\ref{eq:eta}). If a point $q$ does not lie in the convex hull of the curve $\left\{\left(f(x),g(x)\right): a\le x\le b\right\}$, then according to an extension of Carath\'eodory's theorem \cite{Eggleston}, there exist two distinct points $\left(f(X_1),g(X_1)\right)$ and $\left(f(X_2),g(X_2)\right)$, where $X_1,X_2\in[a,b]$, such that the line segment joining them contains the point $q$. Thus, there exists positive numbers $P_1$ and $P_2$, with $P_1+P_2=1$, such that
\[\left|\overline{\Delta L}\right|=c\left|P_1X_1+P_2X_2-\left(P_1X_1^{\gamma}+P_2X_2^{\gamma}\right)^{1/\gamma}\right|.
\]\par
As an example, consider the case $\alpha=2$ or $\gamma=\half$, i.e., $q=2$ and $r=1$. Then, 
\begin{eqnarray*}
|\overline{\Delta L}| &= & c\left|P_1X_1+P_2X_2-\left(P_1X_1^{1/2}+P_2X_2^{1/2}\right)^2\right| \\ & \le & c\left|\half(a+b)-\fourth\left(\sqrt{a}+\sqrt{b}\right)^2\right|\\
& = &c\fourth\left|\left(\sqrt{a}-\sqrt{b}\right)^2\right|=\fourth(z_h-z_c)^2,
\end{eqnarray*}
which, for pure thermal conduction, becomes 
\[\left|\overline{\Delta L}\right|\le \fourth c V^2(t_h-t_c)^2=\fourth c z_h^2\eta_t^2.\]   For processes involving pure deformations, $|\overline{\Delta L}|/L(z_h)\le\fourth\eta_v^2$, which like thermal conduction places the square of the mechanical efficiency, $\eta_v$, as the upper bound on the mean absolute deviation.
\section{Tchebychef's Inequality and Order Statistics}
In treating the irreversible transfer of heat  between any two cells in the system, say $i$ and $j$, what is transferred from $i$, $dQ_i$, is absorbed by $j$, $dQ_j$, or $dQ_i=-dQ_j$. For any pair of heat transfers, there results
\begin{equation}V_i\,dQ_i+V_j\,dQ_j=(V_i-V_j)dQ_i\le0,\label{eq:II-V}
\end{equation}
for the $L$-potential, while
\begin{equation}
\frac{dQ_i}{T_i}+\frac{dQ_j}{T_j}=\left(\frac{1}{T_i}-\frac{1}{T_j}\right)dQ_i\ge0,\label{eq:II-T}
\end{equation}
for the $S$-potential.
Consequently, if there is a transfer of heat from $i\rightarrow j$, then both $V_i<V_j$ and $T_i<T_j$, meaning that the temperatures and volumes of the $n$-cells are similarly ordered.\par
Treating temperature and spatial dimension on the same level, we resort to empirical temperatures, $t_i$, and linear dimensions, $R_i$. The similar ordering of the $n$-tuples, $\vec{t}$ and $\vec{R}$, result in the inequality
\begin{eqnarray}
\mathfrak{M}_r(tR) & = & \left(\sumi p_it_i^rR_i^r\right)^{1/r}\nonumber\\
& \ge & \left(\sumi p_it_i^r\right)^{1/r}
\left(\sumi p_iR_i^r\right)^{1/r} \nonumber\\
& = & \mathfrak{M}_r(t)\mathfrak{M}_r(R), \label{eq:Tchebychef}
\end{eqnarray}
with equality iff all temperatures, $t_i$, and linear dimensions, $R_i$,  are equal.\par Since the mean of order $r$ is the arithmetic mean of $z^r$ raised to the power $1/r$, i.e., $\mathfrak{M}_r(z)=\mathfrak{M}_1(z^r)^{1/r}$, it suffices to consider $r=1$. We then have
\begin{eqnarray*}
\lefteqn{\sumi p_it_iR_i-\sumi p_it_i\,\sumi p_iR_i}\\
& = & \half\sumij\left\{p_ip_jt_jR_j-p_ip_jt_iR_i+p_ip_jt_iR_i-p_ip_jt_jR_i\right\}\\
& = & \half\sumij p_ip_j(t_i-t_j)(R_i-R_j)\ge0,
\end{eqnarray*}
since $\vec{t}$ and $\vec{R}$ are similarly ordered. This is precisely Tchebychef's inequality \cite{HLP}. If $\vec{t}$ and $\vec{R}$ were oppositely ordered then the inequality in (\ref{eq:Tchebychef}) would be reversed, thereby violating the second laws, (\ref{eq:II-V}) and (\ref{eq:II-T}).\par
In other words, the second laws assert that the averages of the product of the temperatures and linear dimensions of the cells, which are in thermal and mechanical contact, cannot be inferior to the product of their means. This is Tchebychef's inequality (\ref{eq:Tchebychef}). Even though these variables have been assumed to be independent, the second laws, (\ref{eq:II-V}) and (\ref{eq:II-T}), introduce correlations through heat transfers by similarly ordering them.\par
\section{From thermodynamics to multifractals and information theory}
The ICG limit also allows a connection to be made with multifractals and information theory \cite{BHL98}, for isothermal processes occurring in nonextensive systems.\par Employing L'H\^opital's rule we get
\begin{equation}
\overline{\Delta S}=\ln\left(\sumi p_iR_i^{D\gamma}\right)-D\gamma\sumi p_i\log R_i, \label{eq:II-bis}
\end{equation}
for the average change in entropy in the ICG limit, where we have set the exponent, $r=D\gamma$. The exponent $D$ is the Hausdorff dimension, defined as
\begin{equation}
\sumi R_i^D=1. \label{eq:Haus}
\end{equation}
Condition (\ref{eq:Haus}) plays a role analogous to the Kraft (in)equality for a uniquely decipherable code.
\par
Considering the exponent $\gamma<1$,  we can apply H\"older's inequality in the reverse form \cite{HLP}
\begin{eqnarray*}
\left[\sumi\left(p^{1/\gamma}_iR_i^D\right)^{\gamma}\right]^{1/\gamma}
\left[\sumi p_i^{-(1/\gamma)\cdot\gamma/(\gamma-1)}\right]^{(\gamma-1)/\gamma}\\
  \le 
\sumi R_i^D=1,
\end{eqnarray*}
on the strength of (\ref{eq:Haus}). With $\gamma=(\alpha-1)/\alpha<1$ and $\alpha>1$, H\"older's inequality becomes
\begin{equation}
\sumi p_iR_i^{D\gamma}\le\left(\sumi p_i^{\alpha}\right)^{1/\alpha}.\label{eq:Holder}
\end{equation}
\par
It is easy to see that we have equality in (\ref{eq:Holder}) iff
\[R_i^D=\frac{p_i^{\alpha}}{\sumi p_i^{\alpha}},\]
or
\begin{equation}
D\ln R_i=\alpha\log p_i-\ln\sumi p_i^{\alpha}, \label{eq:optimal}
\end{equation}
which also satisfies the definition of the Hausdorff dimension, (\ref{eq:Haus}). Multiplying (\ref{eq:optimal}) through by $p_i$, summing, and introducing the result into (\ref{eq:II-bis}) give
\begin{equation}
\overline{\Delta S}=(\alpha-1)\left(S_1-S_{\alpha}\right)>0. \label{eq:S-info}
\end{equation}
The entropies, $S_1$ and $S_\alpha$, are the Shannon-Gibbs,
\[S_1=-\sumi p_i\log p_i,\]
and R\'enyi,
\[S_\alpha=\frac{1}{1-\alpha}\ln\sumi p_i^{\alpha},\]
entropies of order $1$ and $\alpha$, respectively.\par Inequality  (\ref{eq:S-info}) is result  of the fact that for $\alpha>1$, the Shannon entropy is greater than the R\'enyi entropy, while, for $\alpha<1$, the converse is true. This can easily be seen by setting $\beta=\alpha-1$. Then, on the strength of (\ref{eq:mono}) we have
\[\left(\sumi p_ip_i^{\beta}\right)^{1/\beta}>\prodi p_i^{p_i},\]
for $\beta>0$, and the reverse inequality for $\beta<0$. Hence, inequality (\ref{eq:S-info}) is always satisfied. Moreover, in the limit as $\alpha\rightarrow1$, l'H\^opital rule shows that
\[S_1=\lim_{\alpha\rightarrow1}S_{\alpha}=-\sumi p_i\log p_i.\]
In this limit, the average entropy difference, (\ref{eq:S-info}), vanishes.\par Consequently, (\ref{eq:S-info}) shows that in the limit of an isothermal ICG, the average  entropy difference is always proportional to the absolute value of the difference between the Shannon and R\'enyi entropies, when $D$ is identified as the Hausdorff dimension.\par The generalization of the Hausdorff dimension to multifractals, where the generator of cell sizes of lengths $R_i$ with probabilities, $p_i$, require \emph{two\/} exponents \cite{Hal}, 
\begin{equation}\sumi p_i^{\alpha}R_i^{D_{\alpha}(1-\alpha)}=1,\label{eq:Haus-bis}
\end{equation}
where $D_{\alpha}$ is supposed to be some generalization of the Hausdorff dimension, $D$.\par
If $\{p_i\}$ is a complete probability distribution, and $\alpha$ is restricted to the open interval $(0,1)$, in order to ensure that the R\'enyi entropy be concave, then  the usual H\"older inequality, and condition (\ref{eq:Haus-bis}), give
\begin{equation}\sumi R_i^{D_{\alpha}}\ge1.\label{eq:Holder-bis}
\end{equation}
Since for $\alpha=1$, (\ref{eq:optimal}) becomes
\[D_1=\frac{S_1}{\sumi p_i\log(1/R_i)},\]
 it was thought \cite{Hal} that $D_\alpha$ should be related to the R\'enyi entropy in a similar form, viz.,
\[D_{\alpha}=\frac{S_{\alpha}}{\sumi p_i\log(1/R_i)}.\]
\par
This can be derived by averaging
\begin{equation}
D_{\alpha}(1-\alpha)\ln R_i=-\ln\sumi p_i^{\alpha}.\label{eq:D}
\end{equation}
Exponentiating both sides, multiplying by $p_i^{\alpha}$, and summing  do give (\ref{eq:Haus-bis}). But, since the right side of (\ref{eq:D}) is independent of the index $i$, so too must be the left side. This means that all the cell sizes have the same length
\[R^{D_{\alpha}}=\left(\sumi p_i^{\alpha}\right)^{-1/(1-\alpha)}=e^{-S_{\alpha}}.\]
In view of condition (\ref{eq:Holder-bis}), this would imply
\[S_0\ge S_{\alpha}.\]
The entropy is largest in either the state of equal probabilities, or in the state of greatest geometrical regularity. This entropy is the Hartley-Boltzmann entropy, $S_0=\ln n$, which depends on the number of copies considered, and not on their individual frequencies. \par
\section{A measure of the tendency to uniformity}
In this section we show that the internal energy, or the $L$-potential, is a Schur convex function of the empirical temperature and serves as a measure of the tendency of the system to evolve toward a more uniform distribution in temperature.\par
There is no reason to exclude the possibility that the cells in the final state of thermal equilibrium will have different probabilities to be at a given temperature than those in the initial state at the moment the cells have been placed in thermal contact. Let us therefore introduce a second complete set of probabilities $q_1,\ldots,q_n$, which are the probabilities that the final temperatures of the cells will be $\tau_1,\ldots,\tau_n$. If $\tilde{t}$ is any intermediate temperature, we can write the average change in the $L$-potential density, $\ell=L/V$ as
\begin{eqnarray*}
\overline{\Delta\ell} & = & \sumi p_i\int_{t_i}^{\tilde{t}}d\ell(t)+\sumi q_i\int_{\tilde{t}}^{\tau_i}d\ell(t)\\
& = & \sumi q_i\ell(\tau_i)-\sumi p_i\ell(t_i).
\end{eqnarray*}
\par
The process of placing the cells in thermal contact will initiate a process of heat exchange that will ultimately lead to a more uniform temperature distribution. This process of homogenization can be represented mathematically by a doubly stochastic matrix, $\left(\omega_{ij}\right)$, whose rows and columns sum to unity \cite{Mirsky}. \par
The doubly stochastic matrix relates the set of initial temperatures to the set of final temperatures according to
\begin{equation}
\tau_i=\sumj\omega_{ij}t_j, \label{eq:omega}
\end{equation}
for $i=1,\ldots,n$. In other words, the doubly stochastic matrix represents a smoothing operation, and, if the temperatures are ordered in either an increasing or decreasing order, the restriction $\omega_{ij}\in[0,\half]$ will preserve that order.\par
The same doubly stochastic matrix, $(\omega_{ij})$, relates the new probability distribution $\{q_i\}$ to the old one, $\{p_i\}$, in an inverse fashion
\begin{equation}
p_j=\sumi\omega_{ij}q_i, \label{eq:omega-bis}
\end{equation}
for $j=1,\ldots,n$, to that relating the old to the new temperatures, (\ref{eq:omega}).  Multiplying (\ref{eq:omega}) by $q_i$, and summing give
\[\sumi q_i\tau_i=\sumi q_i\sumj\omega_{ij}t_j=\sumj p_jt_j.\]
This says that the system has the same average temperature before and after the cells have been placed in thermal contact. No heat has been transferred between the system and the environment and no work has been performed so that the same average temperature persists.\par
Since $\ell$ is Schur convex,
\begin{equation}
\ell(\tau_i)=\ell\left(\sumi\omega_{ij}t_j\right)\le\sumj\omega_{ij}\ell(t_j),\label{eq:Jensen}
\end{equation}
where the first equality follows from (\ref{eq:omega}). This is none other than Jensen's inequality stating that for a convex function: the function of the average can never be greater than the average of the function.\par Multiplying both sides of (\ref{eq:Jensen}) by $q_i$ and summing result in
\begin{equation}
\sumi q_i\ell(\tau_i)\le\sumi q_i\sumj\omega_{ij}\ell(t_j)=\sumj p_j\ell(t_j), \label{eq:ell-ineq}
\end{equation}
on account of (\ref{eq:omega-bis}). This proves that the $L$-potential  shows a tendency to decrease in the presence of irreversible processes of heat transfer [vid. (\ref{eq:L-ineq})].\par
Inequality (\ref{eq:ell-ineq}) can be most easily established in the case $n=2$. Since the two temperature distributions are related by the doubly stochastic matrix, which in the present case is
\[ \omega=\left(\begin{array}{cc}
\bar{\omega} & \omega\\
\omega & \bar{\omega}
\end{array}\right) \]
where $\bar{\omega}=1-\omega$ for some $\omega\in[0,\half]$, $\vec{t}$ is said to majorize $\vec{\tau}$, or $\vec{t}\succ\vec{\tau}$ \cite{Marshall}.\par  Because $\ell$ is Schur convex
\begin{eqnarray*}
q_1\ell(\tau_1)+q_2\ell(\tau_2) & = & q_1\ell(\bar{\omega}t_1+\omega(t_2)+q_2\ell(\omega t_1+\bar{\omega}t_2)\\
& \le & q_1[\bar{\omega}\ell(t_1)+\omega\ell(t_2)]\\
& + & q_2[\omega\ell(t_1)+\bar{\omega}u(t_2)]\\
& = & p_1\ell(t_1)+p_2\ell(t_2).
\end{eqnarray*}
When the adiabatic walls between the cells have been replaced by diathermal ones, and the system is left to itself, heat will spontaneously flow from the hotter to the colder cells. In economic jargon these flows can be considered as \lq Robin Hood\rq\ operations \cite{Arnold}, where there is a transfer of riches from the wealthy to the poorer segments of the population. This transfer is regulated by the doubly stochastic matrix $\left(\omega_{ij}\right)$.\par
The condition that the final, average temperature persist after thermal contact has been made is the same as saying that the riches have remained the same except they have been spread out more equally. If the gas were ICG, the change in the internal energy would vanish since it is linear in the temperature, and the average temperature has not changed. However, due to the property of Schur convexity of the internal energy, or the $L$-potential, as a function of temperature for an IGG, they show a net tendency to decrease. This property reflects the tendency of the system to reach a more uniform distribution in temperature. In this respect, an IGG offers a more realistic description of Nature than an ICG.

\end{document}